# Generalized Quantum Hydrodynamics and Principles of non-Local Physics

Boris V. Alexeev
Moscow Academy of Fine Chemical Technology (MITHT)
Prospekt Vernadskogo, 86, Moscow 119570, Russia
B.Alexeev@ru.net

**Abstract**

This paper addresses the fundamental principles of generalized Boltzmann physical kinetics, as a part of non-local physics. It is shown that the theory of transport processes (including quantum mechanics) can be considered in the frame of unified theory based on the non-local physical description. Generalized Boltzmann physical kinetics leads to the strict approximation of non-local effects in space and time and after transmission to the local approximation leads to parameter of non-locality $\tau$, which on the quantum level corresponds to the uncertainty principle "time-energy". Schrödinger equation is consequence of the Liouville equation as result of the local approximation of non-local equations. Generalized quantum hydrodynamics leads to Schrödinger equation as a deep particular case of the generalized Boltzmann physical kinetics and therefore of non-local hydrodynamics. Generalized quantum hydrodynamics can be used for solution of fundamental problems in nanoelectronics.

*Key words:* Foundations of the theory of transport processes; The theory of solitons; Generalized hydrodynamic equations; Foundations of quantum mechanics

PACS: 67.55.Fa, 67.55.Hc

**1. Elementary introduction in the basic principles of the Generalized Boltzmann Physical Kinetics.**

In 1872 L Boltzmann published his famous kinetic equation for the one-particle distribution function $f(\mathbf{r}, \mathbf{v}, \mathbf{t})$ [1, 2]. He expressed the equation in the form

$$Df/Dt = J^{st}(f), \qquad (1.1)$$

where $J^{st}$ is the collision integral, and $\frac{D}{Dt} = \frac{\partial}{\partial t} + \mathbf{v} \cdot \frac{\partial}{\partial \mathbf{r}} + \mathbf{F} \cdot \frac{\partial}{\partial \mathbf{v}}$ is the substantial (particle) derivative, $\mathbf{v}$ and $\mathbf{r}$ being the velocity and radius vector of the particle, respectively. Equation (1.1) governs the transport processes in a one-component gas, which is sufficiently rarefied that only binary collisions between particles are of importance and valid only for two character scales, connected with the hydrodynamic time-scale and the time-scale between particle collisions. While we are not concerned here with the explicit form of the collision integral, note that it should satisfy conservation laws of point-like particles in binary collisions. Integrals of the distribution function (i.e. its moments) determine the macroscopic hydrodynamic characteristics of the system, in particular the number density of particles $n$ and the temperature $T$. The Boltzmann equation (BE) is not of course as simple as its symbolic form above might suggest, and it is in only a few special cases that it is amenable to a solution. One example is that of a Maxwellian distribution in a locally, thermodynamically equilibrium gas in the event when no external forces are present. In this case the equality $J^{st} = 0$ and $f = f^{(0)}$ is met, giving the Maxwellian distribution function $f^{(0)}$.

A weak point of the classical Boltzmann kinetic theory is the way it treats the dynamic properties of interacting particles. On the one hand, as the so-called "physical" derivation of the BE suggests, Boltzmann particles are treated as material points; on the other hand, the collision integral in the BE brings into existence the cross sections for collisions between particles. A rigorous approach to the derivation of the kinetic equation for $f$ (noted in following as $(KE_f)$) is based on the hierarchy of the Bogolyubov-Born-Green-Kirkwood-Yvon (BBGKY) [3,4] equations.

A $KE_f$ obtained by the multi-scale method turns into the BE if one ignores the change of the distribution function (DF) over a time of the order of the collision time (or, equivalently, over a length of the order of the particle interaction radius). It is important to note [5-10] that accounting for the third of the scales mentioned above leads (*prior* to introducing any approximation destined to break the Bogolyubov chain) to additional terms, generally of the same order of magnitude, appear in the BE.

If the theory of correlation functions is used to derive $KE_f$ from the BBGKY equations, then the passage to the BE means the neglect of non-local and time delay effects.

Given the above difficulties of the Boltzmann kinetic theory (BKT), the following clearly inter related questions arise:

First, what is a physically infinitesimal volume and how does its introduction (and, as the consequence, the unavoidable smoothing out of the DF) affect the kinetic equation? This question can be formulated in (from the first glance) the paradox form – what is the size of the point in the physical system?

And second, how does a systematic account for the proper diameter of the particle in the derivation of the $KE_f$ affect the Boltzmann equation?

In the theory developed here I shall refer to the corresponding $KE_f$ as the GBE. The derivation of the GBE and the applications of BKT are presented, in particular, in [10]. Accordingly, our purpose is first to explain the essence of the physical generalization of the BE.

Let a particle of finite radius be characterized, as before, by the position vector **r** and velocity **v** of its center of mass at some instant of time $t$. Let us introduce physically small volume (**PhSV**) as element of measurement of macroscopic characteristics of physical system for a point containing in this **PhSV**. We should hope that **PhSV** contains sufficient particles $N_{ph}$ for statistical description of the system and the limit of $N_{ph}$ exists by **PhSV** $\to 0$. In other words all investigating physical system is covering by a net physically small volumes.

Every **PhSV** contains entire quantity of point-like Boltzmann particles, and *the same DF f is prescribed for whole **PhSV** in Boltzmann physical kinetics.* Therefore Boltzmann particles are fully "packed" in the reference volume.

Let us consider two adjoining physically small volumes **PhSV₁** and **PhSV₂**. We have on principle another situation for the particles of finite size (PFS) moving in physical small volumes which are open thermodynamic systems.

Non- local effects take places. Namely:

The fact that center of mass of a PFS is in **PhSV₁** does not mean that entire particle is there. In other words, at any given point in time there are always particles, which are partly inside and partly outside of the reference volume.

*Moreover the particles starting after the last collision near the boundary between two mentioned volumes can change the distribution functions in the neighboring volume. The*

*adjusting of the particles dynamic characteristics for translational degrees of freedom takes several collisions. As result we have in the definite sense "the Knudsen layer" between these volumes. This fact unavoidably leads to fluctuations in mass and hence in other hydrodynamic quantities.*

*Existence of such "Knudsen layers" is not connected with the choice of space nets and fully defined by the reduced description for ensemble of particles of finite diameters in the frame of conception of physically small volumes, therefore – with the chosen method of measurement.*

The animation movie (demonstrating all these effects for particles in the reference volumes) could be proposed to the reader (see my E-mail address), but this movie needs about 38 MB.

The corresponding situation is typical for the theoretical physics – we could remind about the role of probe charge in electrostatics or probe circuit in the physics of magnetic effects.

Suppose that DF $f$ corresponds to **PhSV$_1$** and DF $f - \Delta f$ is connected with **PhSV$_2$** for Boltzmann particles. In the boundary area in the first approximation, fluctuations will be proportional to the mean free path (or, equivalently, to the mean time *between* the collisions). Then for **PhSV** the correction for DF should be introduced as

$$f^a = f - \tau \, Df/Dt \qquad (1.2)$$

in the left hand side of classical BE describing the translation of DF in phase space. As the result

$$Df^a/Dt = J^B, \qquad (1.3)$$

where $J^B$ is the Boltzmann local collision integral.

*Important to notice that it is only qualitative explanation of GBE derivation obtained earlier (see for example [10]) by different strict methods from the BBGKY – chain of kinetic equations.*

The structure of the $KE_f$ is generally as follows

$$\frac{Df}{Dt} = J^B + J^{nl}, \qquad (1.4)$$

where $J^{nl}$ is the non-local integral term incorporating the time delay effect. The generalized Boltzmann physical kinetics, in essence, involves a local approximation

$$J^{nl} = \frac{D}{Dt}\left(\tau \frac{Df_1}{Dt}\right) \qquad (1.5)$$

for the second collision integral, here $\tau$ being the mean time *between* the particle collisions. We can draw here an analogy with the Bhatnagar - Gross - Krook (BGK) approximation for $J^B$,

$$J^B = \frac{f_1^{(0)} - f_1}{\tau}, \qquad (1.6)$$

which popularity as a means to represent the Boltzmann collision integral is due to the huge simplifications it offers.

In other words – the local Boltzmann collision integral admits approximation via the BGK algebraic expression, but more complicated non-local integral can be expressed as differential form (1.5).

The ratio of the second to the first term on the right-hand side of Eqn (1.4) is given to an order of magnitude as $J^{nonlocal}/J^B \approx O(Kn^2)$ and at large Knudsen numbers (defining as ratio of mean free path of particles to the character hydrodynamic length) these terms become of the same order of magnitude. It would seem that at small Knudsen numbers answering to hydrodynamic description the contribution from the second term on the right-hand side of Eqn (1.4) is negligible.

*This is not the case, however.* When one goes over to the hydrodynamic approximation (by multiplying the kinetic equation by collision invariants and then integrating over velocities), the Boltzmann integral part vanishes, and the second term on the right-hand side of Eqn (1.4) gives a single-order contribution in the generalized Navier - Stokes description. Mathematically, we cannot neglect a term with a small parameter in front of the higher derivative. Physically, the appearing additional terms are due to viscosity and they correspond to the small-scale Kolmogorov turbulence [10]. The integral term $J^{nonlocal}$, thus, turns out to be important both at small and large Knudsen numbers in the theory of transport processes. Thus, $\tau Df/Dt$ is the distribution function fluctuation, and writing Eqn (1.3) without taking into account Eqn (1.2) makes the BE non-closed. From the viewpoint of the fluctuation theory, Boltzmann employed the simplest possible closure procedure $f^a = f$. For GBE the generalized H-theorem is proven [7].

Let us consider now some aspects of GBE application beginning with hydrodynamic aspects of the theory. There are two important points to be made here. First, the fluctuations will be proportional to the mean time *between* the collisions (rather than the collision time). This fact is rigorously established in Refs [5-10] and explained above. Therefore, in the first approximation, fluctuations will be proportional to the mean free path $\lambda$ (or, equivalently, to the mean time between the collisions). We can state that the number of particles in reference volume is proportional to cube of the character length $L$ of volume, the number of particles in the surface layer is proportional to $\lambda L^2$, and as result all effect of fluctuation can be estimated as ratio of two mentioned values or as $\lambda/L = Kn$.

Obviously the hydrodynamic equations will explicitly involve fluctuations proportional to $\tau$. For example, the continuity equation changes its form and will contain terms proportional to viscosity. On the other hand - and this is the second point to be made - if the reference volume extends over the whole cavity with the hard walls, then the classical conservation laws should be obeyed, and this is exactly what the paper [8] proves. However, we will here attempt to "guess" the structure of the generalized continuity equation using the arguments outlined above. Neglecting fluctuations, the continuity equation should have the classical form with

$$\rho^a = \rho - \tau A, \quad (\rho \mathbf{v}_0)^a = \rho \mathbf{v}_0 - \tau \mathbf{B},$$

where $\rho$ is density and $\mathbf{v}_0$ is hydrodynamic velocity. Strictly speaking, the factors $A$ and $\mathbf{B}$ can be obtained from the generalized kinetic equation, in our case, from the GBE. Still, we can guess their form without appeal to the $KE_f$.

Indeed, let us write the generalized continuity equation

$$\frac{\partial}{\partial t}(\rho - \tau A) + \frac{\partial}{\partial \mathbf{r}} \cdot (\rho \mathbf{v}_0 - \tau \mathbf{B}) = 0 \qquad (1.7)$$

in the dimensionless form, using $l$, the distance from the reference contour to the hard wall (see Fig. 1.1), as a length scale.

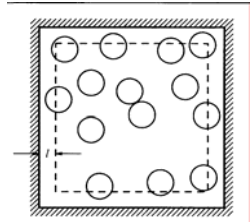

Fig. 1.1. Closed cavity and the reference contour containing particles of a finite diameter.

Then, instead of $\tau$, the (already dimensionless) quantities $A$ and $\mathbf{B}$ will have the Knudsen number $Kn_l = \lambda/l$ as a coefficient. In the limit $l \to 0, Kn_l \to \infty$ the contour embraces the entire cavity contained within hard walls, and there are no fluctuations on the walls. In other words, the classical equations of continuity and motion must be satisfied at the wall. Using hydrodynamic terminology, we note that the conditions $A = 0$, $\mathbf{B} = 0$ correspond to a laminar sub-layer in a turbulent flow. Now if a local Maxwellian distribution is assumed, then the generalized equation of continuity in the Euler approximation is written as

$$\frac{\partial}{\partial t}\left\{\rho - \tau\left[\frac{\partial \rho}{\partial t} + \frac{\partial}{\partial \mathbf{r}}\cdot(\rho\mathbf{v}_0)\right]\right\} + \frac{\partial}{\partial \mathbf{r}}\cdot\left\{\rho\mathbf{v}_0 - \tau\left[\frac{\partial}{\partial t}(\rho\mathbf{v}_0) + \frac{\partial}{\partial \mathbf{r}}\cdot\rho\mathbf{v}_0\mathbf{v}_0 + \vec{I}\cdot\frac{\partial p}{\partial \mathbf{r}} - \rho\mathbf{a}\right]\right\} = 0, \quad (1.8)$$

where $\vec{I}$ is the unit tensor. In the hydrodynamic approximation, the mean time $\tau$ between the collisions is related to the dynamic viscosity $\mu$ by

$$\tau\ p = \Pi\mu, \quad (1.9)$$

where the factor $\Pi$ depends on the choice of a collision model and is $\Pi = 0.8$ for the particular case of neutral gas comprising hard spheres [3]. The generalized hydrodynamic equations (GHE) of energy and motion are much more difficult to guess in this way, making the GBE indispensable.

Now several remarks of principal significance:

1. All fluctuations are found from the strict kinetic considerations and tabulated [10]. The appearing additional terms in GHE are due to viscosity and they correspond to the small-scale Kolmogorov turbulence. The neglect of formally small terms is equivalent, in particular, to dropping the (small-scale) Kolmogorov turbulence from consideration and is the origin of all principal difficulties in usual turbulent theory.

Fluctuations on the wall are equal to zero, from the physical point of view this fact corresponds to laminar sub-layer. Mathematically it leads to additional boundary conditions for GHE. Major difficulties arose when the question of existence and uniqueness of solutions of the Navier - Stokes equations was addressed. O A Ladyzhenskaya has shown for three-dimensional flows that under smooth initial conditions a unique solution is only possible over a finite time interval. Ladyzhenskaya even introduced a "correction" into the Navier - Stokes equations in order that its unique solvability could be proved (see discussion in [11]). GHE do not lead to these difficulties.

2. It would appear that in continuum mechanics the idea of discreteness can be abandoned altogether and the medium under study be considered as a continuum in the literal sense of the word. Such an approach is of course possible and indeed leads to Euler equations in hydrodynamics. But when the viscosity and thermal conductivity effects are to be included, a totally different situation arises. As is well known, the dynamical viscosity is proportional to the mean time $\tau$ between the particle collisions, and a continuum medium in the Euler model with $\tau = 0$ implies that neither viscosity nor thermal conductivity is possible.

3. Many GHE applications were realized for calculation of turbulent flows with the good coincidence with the bench-mark experiments [see for example, 10]. GHE are working with good accuracy even in the theory of sound propagation in the rarefied gases where all moment equations based on the classical BE lead to unsatisfactory results.

4. The non-local kinetic effects listed above will always be relevant to a kinetic theory using one particle description – including, in particular, applications to liquids or plasmas, where self-consistent forces with appropriately cut-off radius of their action are introduced to expand the capability of GBE. The application of the above principles also leads to the modification of the system of the Maxwell electro-dynamic equations (ME). While the traditional formulation of this system does not involve the continuity equation (like (1.7) but for the charge density $\rho^a$ and the current density $\mathbf{j}^a$), nevertheless the ME derivation employs continuity equation and leads to appearance of fluctuations (proportional to $\tau$) of charge density and the current density. In rarefied media both effects lead to Johnson's flicker noise observed in 1925 for the first time by J.B. Johnson by the measurement of current fluctuations of thermo-electron emission. For plasma $\tau$ is the mean time between "close" collisions of charged particles [9, 10].

Finally we can state that introduction of control volume by the reduced description for ensemble of particles of finite diameters leads to fluctuations (proportional to Knudsen number) of velocity moments in the volume. This fact leads to the significant reconstruction of the theory of transport processes. Obviously the mentioned non-local effects can be discussed from positions of breaking of the Bell's inequalities [12] because in the non-local theory the measurement (realized in **PhSV$_1$**) influents on the measurement realized in the adjoining space-time point in **PhSV$_2$** and verse versa.

Generalized quantum hydrodynamics as a part of non-local physics can be used for solution of fundamental problems in nanoelectronics. The review of models and simulation tools based on drift-diffusion model with quantum corrections, kinetic model on solution of Boltzmann equation and quantum hydrodynamics in the frame of local physics can be found in [13].

## 2. Some remarks about hydrodynamic form of Schrödinger equation and its derivation from Liouville equation.

It is well known that basic equation of quantum mechanics – Schrödinger equation (SE) – cannot be strictly derived. SE could be "guessed" from reasonable physical considerations and (after comparison the SE solutions with some experimental data) declared as one of postulates of quantum mechanics. The main steps of this guess can be characterized as follows:

a) The complex function $\psi(x,t)$ is introduced as characteristics of physical objects with corpuscular and wave features. The simplest wave form which could be imagined is:

$$\psi = e^{-i(\omega t - kx)}, \qquad (2.1)$$

with additional conditions

$$\omega = E_\kappa / \hbar, \quad k = 2\pi / \lambda = p / \hbar, \qquad (2.2)$$

where traditional nomenclature is used for frequency $\omega$, kinetic energy $E_k$, wave number $k$ and impulse $p$. Substitution (2.2) in (2.1) and following differentiating (once in time and twice in space) leads to relations

$$i\hbar \frac{\partial \psi}{\partial t} = E_\kappa \psi, \qquad (2.3)$$

$$-\frac{\hbar^2}{2m} \frac{\partial^2 \psi}{\partial x^2} = E_\kappa \psi, \qquad (2.4)$$

because for individual free particle of mass $m$

$$E_\kappa = \frac{p^2}{2m}. \qquad (2.5)$$

As result the one dimensional quantum equation (E. Schrödinger, 1926) takes the form

$$i\hbar \frac{\partial \psi}{\partial t} = -\frac{\hbar^2}{2m} \frac{\partial^2 \psi}{\partial x^2}. \qquad (2.6)$$

b) After obvious generalizations for 3D situation with potential external energy $U(x,y,z,t)$ equation (2.6) takes the form

$$i\hbar \frac{\partial \psi}{\partial t} = -\frac{\hbar^2}{2m} \left( \frac{\partial^2 \psi}{\partial x^2} + \frac{\partial^2 \psi}{\partial y^2} + \frac{\partial^2 \psi}{\partial z^2} \right) + U\psi, \qquad (2.7)$$

as a quantum mechanical postulate. Another manner of differentiating leads to another basic quantum equations (see for example [14]).

The obvious next step should be done and was realized by E. Madelung in 1927 – the derivation of special form of SE after introduction wave function $\psi$ as

$$\psi(x,y,z,t) = \alpha(x,y,z,t) \; e^{i\beta(x,y,z,t)}. \tag{2.8}$$

Using (2.8) and separating the real and imagine parts of Eq. (2.7) one obtains

$$\Delta\alpha - \alpha\left(\frac{\partial\beta}{\partial\mathbf{r}}\right)^2 - \frac{2m}{\hbar^2}\alpha U - \frac{2m}{\hbar}\frac{\partial\beta}{\partial t}\alpha = 0, \tag{2.9}$$

$$\alpha\Delta\beta + 2\frac{\partial\alpha}{\partial\mathbf{r}}\cdot\frac{\partial\beta}{\partial\mathbf{r}} + \frac{2m}{\hbar}\frac{\partial\alpha}{\partial t} = 0. \tag{2.10}$$

But

$$\frac{\partial}{\partial\mathbf{r}}\cdot\left(\alpha^2\frac{\partial\beta}{\partial\mathbf{r}}\right) \equiv \alpha^2\Delta\beta + 2\alpha\frac{\partial\alpha}{\partial\mathbf{r}}\cdot\frac{\partial\beta}{\partial\mathbf{r}}, \tag{2.11}$$

and Eq. (2.10) immediately transforms in continuity equation

$$\frac{\partial\alpha^2}{\partial t} + \frac{\partial}{\partial\mathbf{r}}\cdot\left(\frac{\alpha^2\hbar}{m}\frac{\partial\beta}{\partial\mathbf{r}}\right) = 0, \tag{2.12}$$

if the identifications

$$\rho = \alpha^2, \tag{2.13}$$

$$\mathbf{v} = \frac{\partial}{\partial\mathbf{r}}(\beta\hbar/m) \tag{2.14}$$

introduce in Eq. (2.12). Identification for velocity (2.14) is obvious because for 1D flow

$$v = \frac{\partial}{\partial x}(\beta\hbar/m) = \frac{\hbar}{m}\frac{\partial}{\partial x}\left[-\frac{1}{\hbar}(E_k t - px)\right] = \frac{1}{m}\frac{\partial}{\partial x}(px) = v_\phi,$$

where $v_\phi$ is phase velocity. The existence of the condition (2.14) means that the corresponding flow has potential

$$\varphi = \beta\hbar/m. \tag{2.15}$$

As result two effective hydrodynamic equations take place:

$$\frac{\partial\rho}{\partial t} + \frac{\partial}{\partial\mathbf{r}}\cdot(\rho\mathbf{v}) = 0, \tag{2.16}$$

$$\frac{\partial\mathbf{v}}{\partial t} + \frac{1}{2}\frac{\partial}{\partial\mathbf{r}}v^2 = -\frac{1}{m}\frac{\partial}{\partial\mathbf{r}}\left(U - \frac{\hbar^2}{2m}\frac{\Delta\alpha}{\alpha}\right). \tag{2.17}$$

But

$$\frac{\Delta\alpha}{\alpha} = \frac{\Delta\alpha^2}{2\alpha^2} - \frac{1}{\alpha^2}\left(\frac{\partial\alpha}{\partial\mathbf{r}}\right)^2, \tag{2.18}$$

and the relation (2.18) transforms (2.17) in particular case of the Euler motion equation

$$\frac{\partial\mathbf{v}}{\partial t} + (\mathbf{v}\cdot\frac{\partial}{\partial\mathbf{r}})\mathbf{v} = -\frac{1}{m}\frac{\partial}{\partial\mathbf{r}}U^*, \tag{2.19}$$

where introduced the efficient potential

$$U^* = U - \frac{\hbar^2}{4m\rho}\left[\Delta\rho - \frac{1}{2\rho}\left(\frac{\partial\rho}{\partial\mathbf{r}}\right)^2\right]. \tag{2.20}$$

Additive quantum part of potential can be written in the so called Bohm form

$$\frac{\hbar^2}{2m\sqrt{\rho}}\Delta\sqrt{\rho} = \frac{\hbar^2}{4m\rho}\left[\Delta\rho - \frac{1}{2\rho}\left(\frac{\partial\rho}{\partial\mathbf{r}}\right)^2\right]. \tag{2.21}$$

Then

$$U^* = U + U_{qu} = U - \frac{\hbar^2}{2m\sqrt{\rho}}\Delta\sqrt{\rho} = U - \frac{\hbar^2}{4m\rho}\left[\Delta\rho - \frac{1}{2\rho}\left(\frac{\partial\rho}{\partial\mathbf{r}}\right)^2\right]. \qquad (2.22)$$

Some remarks:
a) SE transforms in hydrodynamic form without additional assumptions. But numerical methods of hydrodynamics are very good developed. As result at the end of seventieth of the last century we realized the systematic calculation of quantum problems using quantum hydrodynamics (see for example [5, 15].
b) SE reduces to the system of continuity equation and particular case of the Euler equation with the additional potential proportional to $\hbar^2$. The physical sense and the origin of the Bohm potential will be established later in this paper.
c) SE (obtained in the frame of the theory of classical complex variables) cannot contain the energy equation in principal. As result in many cases the palliative approach is used when for solution of dissipative quantum problems the classical hydrodynamics is used with insertion of additional Bohm potential in the system of hydrodynamic equations.
d) In spite of Euler form SE leads to reversibility by the change $t \to -t$ only by passing in SE to the complex conjugate values. In other words "the derivation" of SE from (2.1) written as $\psi = e^{i(\omega t - kx)}$ leads to another hydrodynamic equations:

$$-\partial\rho/\partial t + div\ \rho\mathbf{v} = 0, \qquad (2.23)$$

$$-\frac{\partial \mathbf{v}}{\partial t} + (\mathbf{v}\nabla)\mathbf{v} = -\frac{1}{m}\nabla U^*. \qquad (2.24)$$

It means that SE contains in implicit form approximation against of the time arrow. In other words the theory of irreversible processes denies the existence of such processes but the Poincare - Zermelo theorem admits in principal the return in previous state of physical systems obeyed to Newton's dynamics.

Let us consider now from positions of non-local physics the main steps of the SE derivation from Liouville equation. This derivation of classical Schrödinger equation can be found in [16]. Starting point is the Liouville equation written for one particle distribution function $f(x,p,t)$

$$\frac{\partial f}{\partial t} + \frac{p}{m}\frac{\partial f}{\partial x} + F(x)\frac{\partial f}{\partial p} = 0. \qquad (2.25)$$

Eq. (2.25) is collisionless Boltzmann equation, which can't describe dissipation. The external force $F(x) = -\frac{\partial U(x)}{\partial x}$ is acting on the particle with mass $m$. One introduces:
a) the classical probability amplitude $\Psi(x,t)$ for which

$$|\Psi(x,t)|^2 = \int_{-\infty}^{+\infty} f(x,p,t)dp, \qquad (2.26)$$

generally speaking $\Psi(x,t)$ is a complex function;
b) the Wigner type of Fourier transform defined by

$$T[f](x,y,t) = \int_{-\infty}^{+\infty} f(x,p,t)e^{\frac{2ipy}{\alpha}} dp \qquad (2.27)$$

with the a real parameter $\alpha$.

*Transform $T[f](x,y,t)$ introduces the artificial space non-locality in physical system leaving without account the non-locality in time. Transform $T[f](x,y,t)$ has physical meaning if $y \approx 0$.*

In [16] is shown that transform $T[f](x,y,t)$ by $y \neq 0$ can be written as

$$T[f](x,y,t) = \Psi^*(t, x-y)\Psi(t, x+y). \qquad (2.28)$$

The next step is to consider the derivative

$$\frac{\partial}{\partial t}T[f](x,y,t) = \int_{-\infty}^{+\infty} \frac{\partial f(x,p,t)}{\partial t} e^{\frac{2ipy}{\alpha}} dp \qquad (2.29)$$

using the Liouville equation:

$$\frac{\partial}{\partial t}T[f](x,y,t) = -\int_{-\infty}^{+\infty} \left[\frac{p}{m}\frac{\partial f}{\partial x} + F(x)\frac{\partial f}{\partial p}\right] e^{\frac{2ipy}{\alpha}} dp \qquad (2.30)$$

or

$$\frac{\partial}{\partial t}T[f](x,y,t) = \frac{i\alpha}{2m}\int_{-\infty}^{+\infty} \frac{\partial^2}{\partial x \partial y}\left[fe^{\frac{2ipy}{\alpha}}\right] dp + F(x)\frac{2iy}{\alpha}T[f](x,y,t). \qquad (2.31)$$

Eq. (2.31) transforms into

$$i\alpha \frac{\partial}{\partial t}T[f](x,y,t) = -\frac{\alpha^2}{2m}\frac{\partial^2}{\partial x \partial y}T[f](x,y,t) - 2yF(x)T[f](x,y,t) \qquad (2.32)$$

and its solution can be obtained by the perturbation method (see Appendix 1)

$$T[f](t,x,y) = \sum_{n=0}^{\infty} T_n[f](t,x)y^n. \qquad (2.33)$$

Authors [16] transform (2.32) using the substitution:

$$s = x - y, \; r = x + y, \qquad (2.34)$$

$$i\alpha \frac{\partial}{\partial t}T[f](r,s,t) = \left[-\frac{\alpha^2}{2m}\left(\frac{\partial^2}{\partial r^2} - \frac{\partial^2}{\partial s^2}\right) - (r-s)F\left(\frac{r+s}{2}\right)\right]T[f](r,s,t). \qquad (2.35)$$

Using (2.28) in the form

$$T[f](s,r,t) = \Psi^*(t,s)\Psi(t,r), \qquad (2.36)$$

one obtains

$$\Psi^*(t,s)\left[i\alpha \frac{\partial \Psi(t,r)}{\partial t} + \frac{\alpha^2}{2m}\frac{\partial^2 \Psi(t,r)}{\partial r^2} - V(r)\Psi(t,r)\right] =$$
$$\Psi(t,r)\left[-i\alpha \frac{\partial \Psi^*(t,s)}{\partial t} + \frac{\alpha^2}{2m}\frac{\partial^2 \Psi^*(t,s)}{\partial s^2} - V(s)\Psi^*(t,s)\right] \qquad (2.37)$$

After introducing notation

$$K(t,r) = i\alpha \frac{\partial \Psi(t,r)}{\partial t} + \frac{\alpha^2}{2m}\frac{\partial^2 \Psi(t,r)}{\partial r^2} - V(r)\Psi(t,r) \qquad (2.38)$$

Eq. (2.37) is rewriting as

$$\Psi^*(t,s)K(t,r) = \Psi(t,r)K^*(t,s). \qquad (2.39)$$

and can be satisfied identically if

$$K(t,r) = i\alpha \frac{\partial \Psi(t,r)}{\partial t} + \frac{\alpha^2}{2m}\frac{\partial^2 \Psi(t,r)}{\partial r^2} - V(r)\Psi(t,r) = 0. \qquad (2.40)$$

*The assumption* $K(t,r) = i\alpha \frac{\partial \Psi(t,r)}{\partial t} + \frac{\alpha^2}{2m}\frac{\partial^2 \Psi(t,r)}{\partial r^2} - V(r)\Psi(t,r) = 0$ *means from physical point of view the transmission in finalized results to local space approximation of non-local equations.*

Obviously Eq. (2.40) is the Schrödinger equation if arbitrary parameter $\alpha$ coincides with Plank constant $\hbar$ and amplitude $\Psi(x,t)$ transforms into the wave function $\psi(x,t)$.

# 3. Generalized hydrodynamic equations (GHE), quantum hydrodynamics. Schrödinger equation as the consequence of GHE.

The following conclusion of principal significance can be done from the previous consideration:

1. Madelung's quantum hydrodynamics is equivalent to the Schrödinger equation (SE) and leads to description of the quantum particle evolution in the form of Euler equation and continuity equation.
2. SE is consequence of the Liouville equation as result of the local approximation of non-local equations.
3. Generalized Boltzmann physical kinetics leads to the strict approximation of non-local effects in space and time and after transmission to the local approximation leads to parameter $\tau$, which on the quantum level corresponds to the uncertainty principle "time-energy".
4. GHE should lead to SE as a deep particular case of the generalized Boltzmann physical kinetics and therefore of non-local hydrodynamics.

In the following we intend to formulate in explicit form all assumptions which should be done for obtaining of SE from GHE. On the finalized step for simplicity we shall use the non-stationary 1D model without external forces.

Following notations of monograph [10] let us write down equations of the Madelung's hydrodynamics in the form

$$\frac{\partial \rho}{\partial t} + \frac{\partial}{\partial \mathbf{r}} \cdot \rho \mathbf{v}_0 = 0, \tag{3.1}$$

$$\frac{\partial}{\partial t}(\rho \mathbf{v}_0) + \frac{\partial}{\partial \mathbf{r}} \cdot \rho \mathbf{v}_0 \mathbf{v}_0 = -\frac{\rho}{m} \nabla U^*, \tag{3.2}$$

where $\mathbf{v}_0$ is hydrodynamic velocity and potential

$$U^* = U + U_{\kappa\theta} = U - \frac{\hbar^2}{2m\sqrt{\rho}} \Delta\sqrt{\rho} = U - \frac{\hbar^2}{4m\rho}\left[\Delta\rho - \frac{1}{2\rho}\left(\frac{\partial \rho}{\partial \mathbf{r}}\right)^2\right]. \tag{3.3}$$

This form is more convenient for following transformations of GHE.

**Step 1**. Generalized Euler equation can be written in the form [10].

Continuity equation for component $\alpha$:

$$\frac{\partial}{\partial t}\left\{\rho_\alpha - \tau_\alpha^{(0)}\left[\frac{\partial \rho_\alpha}{\partial t} + \frac{\partial}{\partial \mathbf{r}} \cdot (\rho_\alpha \mathbf{v}_0)\right]\right\} +$$

$$+ \frac{\partial}{\partial \mathbf{r}} \cdot \left\{\rho_\alpha \mathbf{v}_0 - \tau_\alpha^{(0)}\left[\frac{\partial}{\partial t}(\rho_\alpha \mathbf{v}_0) + \frac{\partial}{\partial \mathbf{r}} \cdot (\rho_\alpha \mathbf{v}_0 \mathbf{v}_0) + \vec{I} \cdot \frac{\partial p_\alpha}{\partial \mathbf{r}} - \right.\right. \tag{3.4}$$

$$\left.\left. - \rho_\alpha \mathbf{F}_\alpha^{(1)} - \frac{q_\alpha}{m_\alpha} \rho_\alpha \mathbf{v}_0 \times \mathbf{B}\right]\right\} = R_\alpha,$$

Continuity equation for mixture:

$$\frac{\partial}{\partial t}\left\{\rho - \sum_\alpha \tau_\alpha^{(0)}\left[\frac{\partial \rho_\alpha}{\partial t} + \frac{\partial}{\partial \mathbf{r}} \cdot (\rho_\alpha \mathbf{v}_0)\right]\right\} +$$

$$+ \frac{\partial}{\partial \mathbf{r}} \cdot \left\{\rho \mathbf{v}_0 - \sum_\alpha \tau_\alpha^{(0)}\left[\frac{\partial}{\partial t}(\rho_\alpha \mathbf{v}_0) + \frac{\partial}{\partial \mathbf{r}} \cdot (\rho_\alpha \mathbf{v}_0 \mathbf{v}_0) + \vec{I} \cdot \frac{\partial p_\alpha}{\partial \mathbf{r}} - \right.\right. \tag{3.5}$$

$$\left.\left. - \rho_\alpha \mathbf{F}_\alpha^{(1)} - \frac{q_\alpha}{m_\alpha} \rho_\alpha \mathbf{v}_0 \times \mathbf{B}\right]\right\} = 0,$$

Momentum equation

$$\frac{\partial}{\partial t}\left\{\rho_\alpha \mathbf{v}_0 - \tau_\alpha^{(0)}\left[\frac{\partial}{\partial t}(\rho_\alpha \mathbf{v}_0) + \frac{\partial}{\partial \mathbf{r}}\cdot\rho_\alpha \mathbf{v}_0 \mathbf{v}_0 + \frac{\partial p_\alpha}{\partial \mathbf{r}} - \rho_\alpha \mathbf{F}_\alpha^{(1)} - \right.\right.$$

$$\left.\left. -\left(\frac{q_\alpha}{m_\alpha}\right)\rho_\alpha \mathbf{v}_0 \times \mathbf{B}\right]\right\} - \mathbf{F}_\alpha^{(1)}\left[\rho_\alpha - \tau_\alpha^{(0)}\left(\frac{\partial \rho_\alpha}{\partial t} + \frac{\partial}{\partial \mathbf{r}}\cdot(\rho_\alpha \mathbf{v}_0)\right)\right] -$$

$$-\frac{q_\alpha}{m_\alpha}\left\{\rho_\alpha \mathbf{v}_0 - \tau_\alpha^{(0)}\left[\frac{\partial}{\partial t}(\rho_\alpha \mathbf{v}_0) + \frac{\partial}{\partial \mathbf{r}}\cdot\rho_\alpha \mathbf{v}_0 \mathbf{v}_0 + \frac{\partial p_\alpha}{\partial \mathbf{r}} - \rho_\alpha \mathbf{F}_\alpha^{(1)} - \right.\right.$$

$$\left.\left. -\frac{q_\alpha}{m_\alpha}\rho_\alpha \mathbf{v}_0 \times \mathbf{B}\right]\right\} \times \mathbf{B} + \frac{\partial}{\partial \mathbf{r}}\cdot\left\{\rho_\alpha \mathbf{v}_0 \mathbf{v}_0 + p_\alpha \ddot{I} - \tau_\alpha^{(0)}\left[\frac{\partial}{\partial t}(\rho_\alpha \mathbf{v}_0 \mathbf{v}_0 + \right.\right.$$

$$+ p_\alpha \ddot{I}) + \frac{\partial}{\partial \mathbf{r}}\cdot\left(\rho_\alpha (\mathbf{v}_0 \mathbf{v}_0)\mathbf{v}_0 + \rho_\alpha (\mathbf{v}_0 \overline{\mathbf{V}_\alpha})\mathbf{V}_\alpha + \rho_\alpha \overline{(\mathbf{V}_\alpha \mathbf{v}_0)\mathbf{V}_\alpha} + \right.$$

$$+ \rho_\alpha \overline{(\mathbf{V}_\alpha \mathbf{V}_\alpha)}\mathbf{v}_0\big) - \mathbf{F}_\alpha^{(1)}\rho_\alpha \mathbf{v}_0 - \rho_\alpha \mathbf{v}_0 \mathbf{F}_\alpha^{(1)} -$$

$$-\frac{q_\alpha}{m_\alpha}\rho_\alpha[\mathbf{v}_0 \times \mathbf{B}]\mathbf{v}_0 - \frac{q_\alpha}{m_\alpha}\rho_\alpha \overline{[\mathbf{V}_\alpha \times \mathbf{B}]\mathbf{V}_\alpha} -$$

$$\left.\left. -\frac{q_\alpha}{m_\alpha}\rho_\alpha \mathbf{v}_0[\mathbf{v}_0 \times \mathbf{B}] - \frac{q_\alpha}{m_\alpha}\rho_\alpha \overline{\mathbf{V}_\alpha[\mathbf{V}_\alpha \times \mathbf{B}]}\right]\right\} =$$

$$= \int m_\alpha \mathbf{v}_\alpha J_\alpha^{st,el} d\mathbf{v}_\alpha + \int m_\alpha \mathbf{v}_\alpha J_\alpha^{st,inel} d\mathbf{v}_\alpha. \tag{3.6}$$

Energy equation:

$$\frac{\partial}{\partial t}\left\{\frac{\rho v_0^2}{2} + \frac{3}{2}p + \sum_\alpha \varepsilon_\alpha n_\alpha - \sum_\alpha \tau_\alpha^{(0)}\left[\frac{\partial}{\partial t}\left(\frac{\rho_\alpha v_0^2}{2} + \frac{3}{2}p_\alpha + \varepsilon_\alpha n_\alpha\right) + \right.\right.$$

$$\left.\left. +\frac{\partial}{\partial \mathbf{r}}\cdot\left(\frac{1}{2}\rho_\alpha v_0^2 \mathbf{v}_0 + \frac{5}{2}p_\alpha \mathbf{v}_0 + \varepsilon_\alpha n_\alpha \mathbf{v}_0\right) - \mathbf{F}_\alpha^{(1)}\cdot\rho_\alpha \mathbf{v}_0\right]\right\} +$$

$$+\frac{\partial}{\partial \mathbf{r}}\cdot\left\{\frac{1}{2}\rho v_0^2 \mathbf{v}_0 + \frac{5}{2}p\mathbf{v}_0 + \mathbf{v}_0 \sum_\alpha \varepsilon_\alpha n_\alpha - \sum_\alpha \tau_\alpha^{(0)}\left[\frac{\partial}{\partial t}\left(\frac{1}{2}\rho_\alpha v_0^2 \mathbf{v}_0 + \right.\right.\right.$$

$$\left.\left. +\frac{5}{2}p_\alpha \mathbf{v}_0 + \varepsilon_\alpha n_\alpha \mathbf{v}_0\right) + \frac{\partial}{\partial \mathbf{r}}\cdot\left(\frac{1}{2}\rho_\alpha v_0^2 \mathbf{v}_0 \mathbf{v}_0 + \frac{7}{2}p_\alpha \mathbf{v}_0 \mathbf{v}_0 + \frac{1}{2}p_\alpha v_0^2 \ddot{I} + \right.\right.$$

$$\left. +\frac{5}{2}\frac{p_\alpha^2}{\rho_\alpha}\ddot{I} + \varepsilon_\alpha n_\alpha \mathbf{v}_0 \mathbf{v}_0 + \varepsilon_\alpha \frac{p_\alpha}{m_\alpha}\ddot{I}\right) - \rho_\alpha \mathbf{F}_\alpha^{(1)}\cdot \mathbf{v}_0 \mathbf{v}_0 - p_\alpha \mathbf{F}_\alpha^{(1)}\cdot \ddot{I} -$$

$$-\frac{1}{2}\rho_\alpha v_0^2 \mathbf{F}_\alpha^{(1)} - \frac{3}{2}\mathbf{F}_\alpha^{(1)}p_\alpha - \frac{\rho_\alpha v_0^2}{2}\frac{q_\alpha}{m_\alpha}[\mathbf{v}_0 \times \mathbf{B}] - \frac{5}{2}p_\alpha \frac{q_\alpha}{m_\alpha}[\mathbf{v}_0 \times \mathbf{B}] -$$

$$\left. -\varepsilon_\alpha n_\alpha \frac{q_\alpha}{m_\alpha}[\mathbf{v}_0 \times \mathbf{B}] - \varepsilon_\alpha n_\alpha \mathbf{F}_\alpha^{(1)}\right]\right\} - \left\{\mathbf{v}_0 \cdot \sum_\alpha \rho_\alpha \mathbf{F}_\alpha^{(1)} - \sum_\alpha \tau_\alpha^{(0)}\Big[\mathbf{F}_\alpha^{(1)}\cdot\right.$$

$$\left.\cdot\left(\frac{\partial}{\partial t}(\rho_\alpha \mathbf{v}_0) + \frac{\partial}{\partial \mathbf{r}}\cdot\rho_\alpha \mathbf{v}_0 \mathbf{v}_0 + \frac{\partial}{\partial \mathbf{r}}\cdot p_\alpha \ddot{I} - \rho_\alpha \mathbf{F}_\alpha^{(1)} - q_\alpha n_\alpha[\mathbf{v}_0 \times \mathbf{B}]\right)\Big]\right\} = 0. \tag{3.7}$$

Here $\mathbf{F}_\alpha^{(1)}$ are the forces of the non-magnetic origin, $\mathbf{B}$ - magnetic induction, $\ddot{I}$ - unit tensor, $q_\alpha$ - charge of the $\alpha$-component particle, $p_\alpha$ - static pressure for $\alpha$-component, $\mathbf{V}_\alpha$ - thermal velocity, $\varepsilon_\alpha$ - internal energy for the particles of $\alpha$-component.

**Step 2.** One component physical system without external forces.

Continuity equation

$$\frac{\partial}{\partial t}\left\{\rho - \tau\left(\frac{\partial \rho}{\partial t} + \frac{\partial}{\partial \mathbf{r}}\cdot(\rho\mathbf{v}_0)\right)\right\} + \frac{\partial}{\partial \mathbf{r}}\cdot\left\{\rho\mathbf{v}_0 - \tau\left(\frac{\partial}{\partial t}(\rho\mathbf{v}_0) + \right.\right.$$
$$\left.\left. + \frac{\partial}{\partial \mathbf{r}}\cdot(\rho\mathbf{v}_0\mathbf{v}_0) + \frac{\partial p}{\partial \mathbf{r}}\right)\right\} = 0, \quad (3.8)$$

Motion equation

$$\frac{\partial}{\partial t}\left\{\rho v_{0\beta} - \tau\left[\frac{\partial}{\partial t}(\rho v_{0\beta}) + \frac{\partial}{\partial r_\alpha}\left(p\delta_{\alpha\beta} + \rho v^2{}_{0\alpha}\right)\right]\right\} + \frac{\partial}{\partial r_\alpha}\{p\delta_{\alpha\beta} + \rho v_{0\alpha}v_{0\beta} -$$
$$-\tau\left[\frac{\partial}{\partial t}(p\delta_{\alpha\beta} + \rho v_{0\alpha}v_{0\beta}) + \frac{\partial}{\partial r_\gamma}(p\delta_{\alpha\gamma}v_{0\beta} + pv_{0\alpha}\delta_{\beta\gamma} + \right. \quad (3.9)$$
$$\left. + pv_{0\gamma}\delta_{\alpha\beta} + \rho v_{0\alpha}v_{0\beta}v_{0\gamma})\right]\} = 0,$$

In Eq. (3.9) Einstein's rule is used for summation with index $\alpha, \beta, \gamma = 1,2,3$.

Energy equation

$$\frac{\partial}{\partial t}\left\{3p + \rho v_0^2 - \tau\left[\frac{\partial}{\partial t}(3p + \rho v_0^2) + \frac{\partial}{\partial \mathbf{r}}\cdot(\mathbf{v}_0(\rho v_0^2 + 5p))\right]\right\} +$$
$$+ \frac{\partial}{\partial \mathbf{r}}\cdot\left\{\begin{array}{l}\mathbf{v}_0(\rho v_0^2 + 5p) - \\ -\tau\left[\frac{\partial}{\partial t}(\mathbf{v}_0(\rho v_0^2 + 5p)) + \frac{\partial}{\partial \mathbf{r}}\cdot\left(\ddot{\mathrm{I}}pv_0^2 + \rho v_0^2\mathbf{v}_0\mathbf{v}_0 + 7p\mathbf{v}_0\mathbf{v}_0 + 5\ddot{\mathrm{I}}\frac{p^2}{\rho}\right)\right]\end{array}\right\} = 0.$$
$$(3.10)$$

**Step 3.** Transmission to the non-stationary 1D model for the generalized Euler equations.
By the following inscription $\tau = \tau^{(qu)}$ corresponds to choice of scale on the quantum level.

Continuity equation:

$$\frac{\partial}{\partial t}\left\{\rho - \tau^{(qu)}\left[\frac{\partial \rho}{\partial t} + \frac{\partial}{\partial x}(\rho v_0)\right]\right\} + \frac{\partial}{\partial x}\left\{\rho v_0 - \tau^{(qu)}\left[\frac{\partial}{\partial t}(\rho v_0) + \right.\right.$$
$$\left.\left. + \frac{\partial}{\partial x}(\rho v_0^2) + \frac{\partial p}{\partial x}\right]\right\} = 0, \quad (3.11)$$

Motion equation:

$$\frac{\partial}{\partial t}\left\{\rho v_0 - \tau^{(qu)}\left[\frac{\partial}{\partial t}(\rho v_0) + \frac{\partial}{\partial x}(\rho v_0^2) + \frac{\partial p}{\partial x}\right]\right\} +$$
$$+ \frac{\partial}{\partial x}\left\{\rho v_0^2 + p - \tau^{(qu)}\left[\frac{\partial}{\partial t}(\rho v_0^2 + p) + \frac{\partial}{\partial x}(\rho v_0^3 + 3pv_0)\right]\right\} = 0, \quad (3.12)$$

Energy equation:

$$\frac{\partial}{\partial t}\left\{\rho v_0^2 + 3p - \tau^{(qu)}\left[\frac{\partial}{\partial t}\left(\rho v_0^2 + 3p\right) + \frac{\partial}{\partial x}\left(\rho v_0^3 + 5pv_0\right)\right]\right\} +$$

$$+ \frac{\partial}{\partial x}\left\{\rho v_0^3 + 5pv_0 - \tau^{(qu)}\left[\frac{\partial}{\partial t}\left(\rho v_0^3 + 5pv_0\right) + \frac{\partial}{\partial x}\left(\rho v_0^4 +\right.\right.\right. \quad (3.13)$$

$$\left.\left.\left.+ 8pv_0^2 + 5\frac{p^2}{\rho}\right)\right]\right\} = 0.$$

**Step 4.** All the time non-local terms are omitted following the Schrödinger-Madelung model.

Continuity equation:

$$\frac{\partial \rho}{\partial t} + \frac{\partial}{\partial x}\left\{\rho v_0 - \tau^{(qu)}\left[\frac{\partial}{\partial t}(\rho v_0) +\right.\right.$$

$$\left.\left.+ \frac{\partial}{\partial x}(\rho v_0^2) + \frac{\partial p}{\partial x}\right]\right\} = 0, \quad (3.14)$$

Momentum equation:

$$\frac{\partial}{\partial t}\{\rho v_0\} +$$

$$+ \frac{\partial}{\partial x}\left\{\rho v_0^2 + p - \tau^{(qu)}\left[\frac{\partial}{\partial t}\left(\rho v_0^2 + p\right) + \frac{\partial}{\partial x}\left(\rho v_0^3 + 3pv_0\right)\right]\right\} = 0, \quad (3.15)$$

Energy equation:

$$\frac{\partial}{\partial t}\left\{\rho v_0^2 + 3p\right\} +$$

$$+ \frac{\partial}{\partial x}\left\{\rho v_0^3 + 5pv_0 - \tau^{(qu)}\left[\frac{\partial}{\partial t}\left(\rho v_0^3 + 5pv_0\right) + \frac{\partial}{\partial x}\left(\rho v_0^4 +\right.\right.\right. \quad (3.16)$$

$$\left.\left.\left.+ 8pv_0^2 + 5\frac{p^2}{\rho}\right)\right]\right\} = 0.$$

**Step 5.** All terms containing the static pressure are omitted following the Schrödinger-Madelung model.

Continuity equation:

$$\frac{\partial \rho}{\partial t} + \frac{\partial}{\partial x}\left\{\left\{\rho v_0 - \tau^{(qu)}\left[\frac{\partial}{\partial t}(\rho v_0) + \frac{\partial}{\partial x}(\rho v_0^2)\right]\right\}\right\} = 0, \quad (3.17)$$

Momentum equation:

$$\frac{\partial}{\partial t}\{\rho v_0\} + \frac{\partial}{\partial x}\left\{\rho v_0^2 - \tau^{(qu)}\left[\frac{\partial}{\partial t}(\rho v_0^2) + \frac{\partial}{\partial x}(\rho v_0^3)\right]\right\} = 0, \quad (3.18)$$

Energy equation:

$$\frac{\partial}{\partial t}\{\rho v_0^2\} + \frac{\partial}{\partial x}\left\{\rho v_0^3 - \tau^{(qu)}\left[\frac{\partial}{\partial t}(\rho v_0^3) + \frac{\partial}{\partial x}(\rho v_0^4)\right]\right\} = 0. \quad (3.19)$$

The following simplification of this system of equations will be done after estimation of non-local parameter $\tau^{(qu)}$ from the Heisenberg uncertainty principle which can be written as

$$\sqrt{\overline{(\Delta x)^2}}\sqrt{\overline{(\Delta p_x)^2}} \geq \hbar/2, \qquad (3.20)$$

or

$$\sqrt{\overline{p_x^2}}\sqrt{\overline{x^2}} \geq \hbar/2. \qquad (3.21)$$

We use the estimate

$$mux \cong \hbar/2, \quad \frac{mu^2}{2}\frac{x}{u} \cong \frac{\hbar}{4}, \quad E\tau^{(qu)} \cong \frac{\hbar}{4}, \quad \tau^{(qu)} \cong \frac{1}{4\omega} = \frac{1}{4uk} = \frac{\lambda}{8\pi u}, \qquad (3.22)$$

where $u$ is velocity along the $x$ axes. Now we formulate the next step.

**Step 6**. Following the Schrödinger - Madelung model we omit all explicit time dependence of non-local terms.
Continuity equation:

$$\frac{\partial \rho}{\partial t} + \frac{\partial}{\partial x}\left[\rho u - \tau^{(qu)}\frac{\partial}{\partial x}\left(\rho u^2\right)\right] = 0, \qquad (3.23)$$

Motion equation:

$$\frac{\partial}{\partial t}(\rho u) + \frac{\partial}{\partial x}\left[\rho u^2 - \tau^{(qu)}\frac{\partial}{\partial x}\left(\rho u^3\right)\right] = 0, \qquad (3.24)$$

Energy equation:

$$\frac{\partial}{\partial t}\left(\rho u^2\right) + \frac{\partial}{\partial x}\left[\rho u^3 - \tau^{(qu)}\frac{\partial}{\partial x}\left(\rho u^4\right)\right] = 0. \qquad (3.25)$$

On this step we write down the Madelung equations in chosen nomenclature and assumptions.

$$\frac{\partial \rho}{\partial t} + \frac{\partial}{\partial x}(\rho u) = 0, \qquad (3.26)$$

$$\frac{\partial}{\partial t}(\rho u) + \frac{\partial}{\partial x}\left(\rho u^2\right) = \frac{\hbar^2}{4m^2}\rho\frac{\partial}{\partial x}\left\{\frac{1}{\rho}\left[\frac{\partial^2 \rho}{\partial x^2} - \frac{1}{2\rho}\left(\frac{\partial \rho}{\partial x}\right)^2\right]\right\}, \qquad (3.27)$$

**Step 7.** Following the Schrödinger - Madelung model we reduce the system of quantum hydrodynamic equations to the continuity and motion equations.
This assumption leads to the condition following from the energy equation (3.19).

$$\rho u^3 = \tau^{(qu)}\frac{\partial}{\partial x}\left(\rho u^4\right). \qquad (3.28)$$

In this case the energy equation (3.25) can be transformed in the relation

$$\frac{\partial}{\partial t}\left(\rho u^2\right) = 0. \qquad (3.29)$$

or

$$\rho u^2 = C(x). \qquad (3.30)$$

This energy conservation law $\rho u^2 = C(x)$ does not lead out of the limits of approximation for the formulated assumptions. Moreover this energy space dependence remains in Eqs. (3.23), (3.24).

*It is shown below that step 7 and condition (3.28) lead to the Bohm potential.*
After substitution of the condition (3.28) in Eq. (3.24) we reach the system of two hydrodynamic equations

$$\frac{\partial \rho}{\partial t} + \frac{\partial}{\partial x}\left[\rho u - \tau^{(qu)}\frac{\partial}{\partial x}(\rho u^2)\right] = 0, \qquad (3.31)$$

$$\frac{\partial}{\partial t}(\rho u) + \frac{\partial}{\partial x}(\rho u^2) = \frac{\partial}{\partial x}\left\{\tau^{(qu)}\frac{\partial}{\partial x}\left[\tau^{(qu)}\frac{\partial}{\partial x}(\rho u^4)\right]\right\}, \qquad (3.32)$$

The motion equation (3.32) can be written as

$$\frac{\partial}{\partial t}(\rho u) + \frac{\partial}{\partial x}(\rho u^2) = \frac{\partial}{\partial x}\left\{\tau^{(qu)2}\frac{\partial^2}{\partial x^2}(\rho u^4) + \tau^{(qu)}\frac{\partial \tau^{(qu)}}{\partial x}\frac{\partial}{\partial x}(\rho u^4)\right\}, \qquad (3.33)$$

**Step 8.** Following the Schrödinger - Madelung model we omit non-local terms in continuity equation.

As result we have the continuity equation coinciding with the continuity Madelung's equation

$$\frac{\partial \rho}{\partial t} + \frac{\partial}{\partial x}(\rho u) = 0, \qquad (3.34)$$

Let us transform now the derivative

$$\frac{\partial^2}{\partial x^2}(\rho u^4) = u^4 \frac{\partial^2 \rho}{\partial x^2} + 8u^3 \frac{\partial u}{\partial x}\frac{\partial \rho}{\partial x} + 12 u^2 \rho \left(\frac{\partial u}{\partial x}\right)^2 + 4u^3 \rho \frac{\partial^2 u}{\partial x^2} \qquad (3.35)$$

**Step 9.** One conserves in the motion equation only terms proportional to the senior powers of velocity.

After substitution of (3.35) in (3.33) one obtains

$$\frac{\partial}{\partial t}(\rho u) + \frac{\partial}{\partial x}(\rho u^2) = \frac{\partial}{\partial x}\left\{\tau^{(qu)2}\left[u^4 \frac{\partial^2 \rho}{\partial x^2} + 8u^3 \frac{\partial u}{\partial x}\frac{\partial \rho}{\partial x}\right]\right\} + \frac{\partial}{\partial x}\left[\tau^{(qu)}\frac{\partial \tau^{(qu)}}{\partial x}\frac{\partial}{\partial x}(\rho u^4)\right]$$
.
$$(3.36)$$

**Step 10.** One introduces the estimate from continuity equation (3.34) for the quasi-stationary case.

$$\frac{\partial}{\partial x}(\rho u) = 0, \quad u\frac{\partial \rho}{\partial x} = -\rho\frac{\partial u}{\partial x}, \quad \frac{1}{u}\frac{\partial u}{\partial x} = -\frac{1}{\rho}\frac{\partial \rho}{\partial x}, \qquad (3.37)$$

As result from (3.36), (3.37):

$$\frac{\partial}{\partial t}(\rho u) + \frac{\partial}{\partial x}(\rho u^2) = \frac{\partial}{\partial x}\left\{\tau^{(qu)2} u^4 \left[\frac{\partial^2 \rho}{\partial x^2} - 8\frac{1}{\rho}\left(\frac{\partial \rho}{\partial x}\right)^2\right]\right\} + \frac{\partial}{\partial x}\left\{\tau^{(qu)}\frac{\partial \tau^{(qu)}}{\partial x}\frac{\partial}{\partial x}(\rho u^4)\right\}$$
$$(3.38)$$

**Step 11.** One neglects of the $\tau^{(qu)}$ dependence in space.
One obtains from Eq. (3.38)

$$\frac{\partial}{\partial t}(\rho u) + \frac{\partial}{\partial x}(\rho u^2) = \frac{\partial}{\partial x}\left\{\tau^{(qu)2} u^4 \left[\frac{\partial^2 \rho}{\partial x^2} - 8\frac{1}{\rho}\left(\frac{\partial \rho}{\partial x}\right)^2\right]\right\}. \qquad (3.39)$$

**Step 12.** Introduction of estimates for $\tau^{(qu)}$ in the explicit form in the motion equation.
Using (3.22) and de Broglie relation $p = h/\lambda$, we find

$$\tau^{(qu)} \cong \frac{1}{4\omega} = \frac{1}{4uk} = \frac{\lambda}{8\pi u},$$

$$\tau^{(qu)2} u^4 = \left(\frac{\lambda}{8\pi u}\right)^2 u^4 = \left(\frac{\lambda u}{8\pi}\right)^2 = \left(\frac{\lambda p}{8\pi m}\right)^2 = \left(\frac{h}{8\pi m}\right)^2 = \left(\frac{\hbar}{4m}\right)^2. \quad (3.40)$$

We have from (3.39), (3.40):

$$\frac{\partial}{\partial t}(\rho u) + \frac{\partial}{\partial x}(\rho u^2) = \frac{\partial}{\partial x}\left\{\frac{\hbar^2}{16m^2}\left[\frac{\partial^2 \rho}{\partial x^2} - 8\frac{1}{\rho}\left(\frac{\partial \rho}{\partial x}\right)^2\right]\right\}. \quad (3.41)$$

Rewrite (3.41) as follows

$$\frac{\partial}{\partial t}(\rho u) + \frac{\partial}{\partial x}(\rho u^2) = \frac{\hbar^2}{16m^2}\frac{\partial}{\partial x}\left\{\rho\frac{1}{\rho}\left[\frac{\partial^2 \rho}{\partial x^2} - 8\frac{1}{\rho}\left(\frac{\partial \rho}{\partial x}\right)^2\right]\right\} \quad (3.42)$$

and after obvious differentiation in (3.42)

$$\frac{\partial}{\partial t}(\rho u) + \frac{\partial}{\partial x}(\rho u^2) = \rho\frac{\partial}{\partial x}\left\{\frac{1}{\rho}\frac{\hbar^2}{16m^2}\left[\frac{\partial^2 \rho}{\partial x^2} - 8\frac{1}{\rho}\left(\frac{\partial \rho}{\partial x}\right)^2\right]\right\} +$$
$$+ \frac{1}{\rho}\frac{\partial \rho}{\partial x}\frac{\hbar^2}{16m^2}\left[\frac{\partial^2 \rho}{\partial x^2} - 8\frac{1}{\rho}\left(\frac{\partial \rho}{\partial x}\right)^2\right] \quad (3.43)$$

The last term in (3.43) can be neglected. Really we have

$$\frac{\partial}{\partial t}(\rho u) + \frac{\partial}{\partial x}(\rho u^2) = \rho\frac{\hbar^2}{4m^2}\frac{\partial}{\partial x}\left\{\frac{1}{\rho}\left[0.25\frac{\partial^2 \rho}{\partial x^2} - 2\frac{1}{\rho}\left(\frac{\partial \rho}{\partial x}\right)^2\right]\right\} +$$
$$+ \frac{\hbar^2}{16m^2}\frac{\partial \rho}{\partial x}\frac{\partial \ln \rho}{\partial x}\left[\frac{\partial}{\partial x}\left(\ln\frac{\partial \rho}{\partial x}\right) - 8\left(\frac{\partial \ln \rho}{\partial x}\right)\right] \quad (3.44)$$

and neglecting the derivatives of the logarithmic terms one obtains

$$\frac{\partial}{\partial t}(\rho u) + \frac{\partial}{\partial x}(\rho u^2) = \frac{\hbar^2}{4m^2}\rho\frac{\partial}{\partial x}\left\{\frac{1}{\rho}\left[0.25\frac{\partial^2 \rho}{\partial x^2} - 2\frac{1}{\rho}\left(\frac{\partial \rho}{\partial x}\right)^2\right]\right\}. \quad (3.45)$$

Let us compare now Eq. (3.45), obtained from the non-local hydrodynamic with Madelung equation rewritten here ones more:

$$\frac{\partial}{\partial t}(\rho u) + \frac{\partial}{\partial x}(\rho u^2) = \frac{\hbar^2}{4m^2}\rho\frac{\partial}{\partial x}\left\{\frac{1}{\rho}\left[\frac{\partial^2 \rho}{\partial x^2} - \frac{1}{2\rho}\left(\frac{\partial \rho}{\partial x}\right)^2\right]\right\}. \quad (3.46)$$

Equations (3.45) and (3.46) can be written in the unified form

$$\frac{\partial}{\partial t}(\rho u) + \frac{\partial}{\partial x}(\rho u^2) = \frac{\hbar^2}{4m^2}\rho\frac{\partial}{\partial x}\left\{\frac{1}{\rho}\left[\gamma\frac{\partial^2 \rho}{\partial x^2} - \delta\frac{1}{\rho}\left(\frac{\partial \rho}{\partial x}\right)^2\right]\right\}, \quad (3.47)$$

the numerical coefficients $\gamma = 1$, $\delta = 0.5$ correspond to Schrödinger equation.

Therefore the condition (3.28) $\rho u^3 = \tau^{(qu)}\frac{\partial}{\partial x}(\rho u^4)$ leads by $\gamma = 1$, $\delta = 0.5$ to the Bohm potential, reflecting "the last traces" of omitted energy equation. In the other words Bohm potential reflects the condition of the dissipation absence in the particular case, when $\gamma = 1$, $\delta = 0.5$.

Rejection of some assumptions formulated above could change the coefficients $\gamma, \delta$ in (3.47). Let us consider for example the possible coordinate dependence of the non-local value $\tau^{(qu)}$. Substituting (3.40) in (3.38) we find after differentiation without taking into account the possible space dependence of the particle wave length

$$\frac{\partial}{\partial t}(\rho u) + \frac{\partial}{\partial x}(\rho u^2) = \frac{\partial}{\partial x}\left\{ \frac{\hbar^2}{16 m^2}\left[ \frac{\partial^2 \rho}{\partial x^2} - 8\frac{1}{\rho}\left(\frac{\partial \rho}{\partial x}\right)^2 \right]\right\} - \frac{\partial}{\partial x}\left\{ \tau^{(\kappa s)} \frac{\lambda}{8\pi u^2} \frac{\partial u}{\partial x}\frac{\partial}{\partial x}(\rho u^4)\right\}$$
(3.48)

or

$$\frac{\partial}{\partial t}(\rho u) + \frac{\partial}{\partial x}(\rho u^2) = \frac{\partial}{\partial x}\left\{ \frac{\hbar^2}{16 m^2}\left[ \frac{\partial^2 \rho}{\partial x^2} - 8\frac{1}{\rho}\left(\frac{\partial \rho}{\partial x}\right)^2 \right]\right\} - \frac{\partial}{\partial x}\left\{ \tau^{(\kappa s)2} \frac{1}{u} \frac{\partial u}{\partial x}\frac{\partial}{\partial x}(\rho u^4)\right\}$$
(3.49)

After transformations we find the analogue equation (see (3.45)) but with another numerical coefficients $\gamma, \delta$.

$$\frac{\partial}{\partial t}(\rho u) + \frac{\partial}{\partial x}(\rho u^2) = \frac{\hbar^2}{4 m^2}\frac{\partial}{\partial x}\left\{\left[ 0.25\frac{\partial^2 \rho}{\partial x^2} - 2.75\frac{1}{\rho}\left(\frac{\partial \rho}{\partial x}\right)^2 \right]\right\}. \qquad (3.50)$$

**Conclusion: Schrödinger equation is a deep particular case of the generalized hydrodynamic equations.**

### 4. Some another applications of the generalized hydrodynamic equations (GHE).

Generalized hydrodynamic equations admit the solution of the physical problems in different scientific fields in the frame of unified theory. Many examples of this kind (including strict theory of turbulent flows) can be found in monograph [10]. Let us deliver some examples on the basement written above with the aim to derive SE from GHE.

**A. To the solitons theory.**

Let us consider equations (3.31), (3.32) obtained above after the step 7.

$$\frac{\partial \rho}{\partial t} + \frac{\partial}{\partial x}\left[ \rho u - \tau^{(qu)}\frac{\partial}{\partial x}(\rho u^2) \right] = 0, \qquad (4.1)$$

$$\frac{\partial}{\partial t}(\rho u) + \frac{\partial}{\partial x}(\rho u^2) = \frac{\partial}{\partial x}\left\{ \tau^{(qu)}\frac{\partial}{\partial x}\left[ \tau^{(qu)}\frac{\partial}{\partial x}(\rho u^4) \right]\right\}, \qquad (4.2)$$

Suppose that $\tau^{(qu)} = const$ and introduce the following scale system:
$\rho_0$, $u_0$, $t_0 = \tau^{(\kappa s)}$, $x_0 = u_0 t_0$. In dimensionless form Eqs. (4.1), (4.2) take the form:

continuity equation

$$\frac{\partial \tilde{\rho}}{\partial \tilde{t}} + \frac{\partial}{\partial \tilde{x}}\left[ \tilde{\rho}\tilde{u} - \frac{\partial}{\partial \tilde{x}}(\tilde{\rho}\tilde{u}^2) \right] = 0, \qquad (4.3)$$

motion equation

$$\frac{\partial}{\partial \tilde{t}}(\tilde{\rho}\tilde{u}) + \frac{\partial}{\partial \tilde{x}}(\tilde{\rho}\tilde{u}^2) = \frac{\partial^3}{\partial \tilde{x}^3}(\tilde{\rho}\tilde{u}^4). \qquad (4.4)$$

With the aim to find wave solutions of Eqs. (4.3), (4.4) we use moving coordinate system, where
$$\tilde{\xi} = \tilde{x} + \tilde{C}\tilde{t}. \tag{4.5}$$
In this new coordinate system $\tilde{\rho} = \tilde{\rho}(\tilde{\xi}, \tilde{t})$, $\tilde{u} = \tilde{u}(\tilde{\xi}, \tilde{t})$ and for these variables the system takes the form

$$\tilde{C}\frac{\partial \tilde{\rho}}{\partial \tilde{\xi}} + \frac{\partial \tilde{\rho}}{\partial \tilde{t}} + \frac{\partial}{\partial \tilde{\xi}}\left[\tilde{\rho}\tilde{u} - \frac{\partial}{\partial \tilde{\xi}}(\tilde{\rho}\tilde{u}^2)\right] = 0, \tag{4.6}$$

$$\tilde{C}\frac{\partial}{\partial \tilde{\xi}}(\tilde{\rho}\tilde{u}) + \frac{\partial}{\partial \tilde{t}}(\tilde{\rho}\tilde{u}) + \frac{\partial}{\partial \tilde{\xi}}(\tilde{\rho}\tilde{u}^2) = \frac{\partial^3}{\partial \tilde{\xi}^3}(\tilde{\rho}\tilde{u}^4). \tag{4.7}$$

For the soliton solution there is no explicit dependence on time for coordinate system moving with the phase velocity $\tilde{C}$. We have

$$\tilde{C}\frac{\partial \tilde{\rho}}{\partial \tilde{\xi}} + \frac{\partial}{\partial \tilde{\xi}}\left[\tilde{\rho}\tilde{u} - \frac{\partial}{\partial \tilde{\xi}}(\tilde{\rho}\tilde{u}^2)\right] = 0, \tag{4.8}$$

$$\tilde{C}\frac{\partial}{\partial \tilde{\xi}}(\tilde{\rho}\tilde{u}) + \frac{\partial}{\partial \tilde{\xi}}(\tilde{\rho}\tilde{u}^2) = \frac{\partial^3}{\partial \tilde{\xi}^3}(\tilde{\rho}\tilde{u}^4). \tag{4.9}$$

After the first integration:
$$(\tilde{u} + \tilde{C})\tilde{\rho} = \frac{\partial}{\partial \tilde{\xi}}(\tilde{\rho}\tilde{u}^2) + \tilde{C}_1. \tag{4.10}$$

$$(\tilde{u} + \tilde{C})\tilde{\rho}\tilde{u} = \frac{\partial^2}{\partial \tilde{\xi}^2}(\tilde{\rho}\tilde{u}^4) + \tilde{C}_2. \tag{4.11}$$

From the conservation laws follow that constants $\tilde{C}_1$ and $\tilde{C}_2$ can be chosen equal to zero. Then

$$\frac{\partial}{\partial \tilde{\xi}}(\tilde{\rho}\tilde{u}^2) = (\tilde{u} + \tilde{C})\tilde{\rho}, \tag{4.12}$$

$$\frac{\partial^2}{\partial \tilde{\xi}^2}(\tilde{\rho}\tilde{u}^4) = (\tilde{u} + \tilde{C})\tilde{\rho}\tilde{u}. \tag{4.13}$$

Differential equation (4.12) can be written in the form
$$\frac{\partial}{\partial \tilde{\xi}}\ln \tilde{\rho}\tilde{u}^2 = \frac{\tilde{u} + \tilde{C}}{\tilde{u}^2}, \tag{4.14}$$

and has the following solution:
$$\tilde{\rho}\tilde{u}^2 = \tilde{C}_3 e^{\int_0^{\xi}\frac{\tilde{u}+\tilde{C}}{\tilde{u}^2}d\xi}. \tag{4.15}$$

Let us transform (4.13) substituting the solution (4.15) in Eq. (4.13). This substitution leads to exclusion of the variable $\tilde{\rho}$. As result we have

$$2\tilde{u}^2\frac{\partial}{\partial \tilde{\xi}}\left(\tilde{u}\frac{\partial \tilde{u}}{\partial \xi}\right) + \left(3\tilde{u}^2 + 2\tilde{C}\tilde{u}\right)\frac{\partial \tilde{u}}{\partial \tilde{\xi}} + \tilde{C}(\tilde{u} + \tilde{C}) = 0. \tag{4.16}$$

The system of stiff equations (4.12), (4.16) can be solved by numerical methods. Let us show some results of such integration. In all cases three initial conditions were introduced; Fig. 4.1 corresponds to conditions: $\tilde{u}(0) = 1$, $\frac{\partial \tilde{u}}{\partial \tilde{\xi}}(0) = -1$, $\tilde{\rho}(0) = 1$; Fig. 4.2 corresponds to conditions: $\tilde{u}(0) = 1$, $\frac{\partial \tilde{u}}{\partial \tilde{\xi}}(0) = 1$, $\tilde{\rho}(0) = 1$; Fig. 4.3 corresponds to conditions: $\tilde{u}(100) = 1$, $\frac{\partial \tilde{u}}{\partial \tilde{\xi}}(100) = -1$, $\tilde{\rho}(100)=1$; Fig. 4.4 corresponds to conditions: $\tilde{u}(100)=1$, $\dfrac{\partial \tilde{u}}{\partial \tilde{\xi}}(100)=1$, $\tilde{\rho}(100)=1$. In all cases dimensionless phase velocity $\tilde{C}=1$.

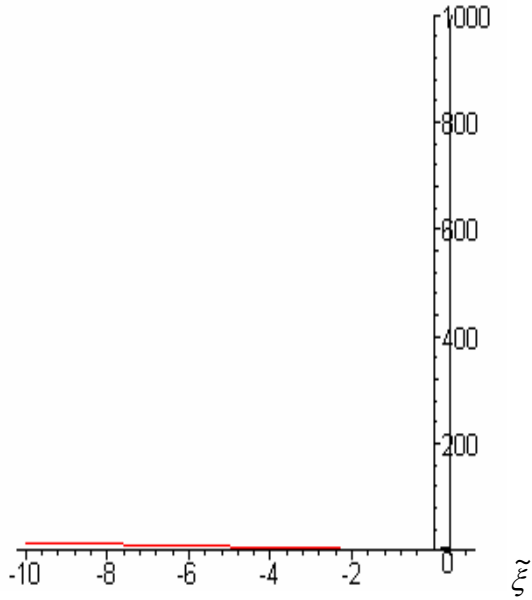
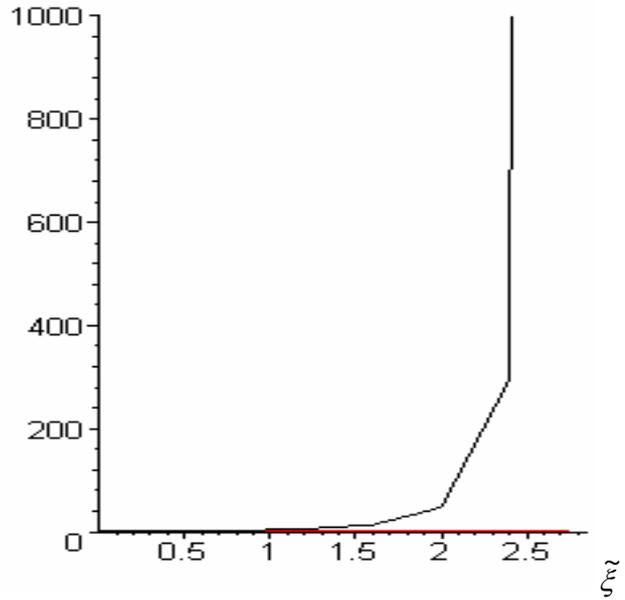

Fig. 4.1. $\tilde{u}(0)=1, \dfrac{\partial \tilde{u}}{\partial \tilde{\xi}}(0)=-1, \tilde{\rho}(0)=1$    Fig. 4.2. $\tilde{u}(0)=1, \dfrac{\partial \tilde{u}}{\partial \tilde{\xi}}(0)=1, \tilde{\rho}(0)=1$

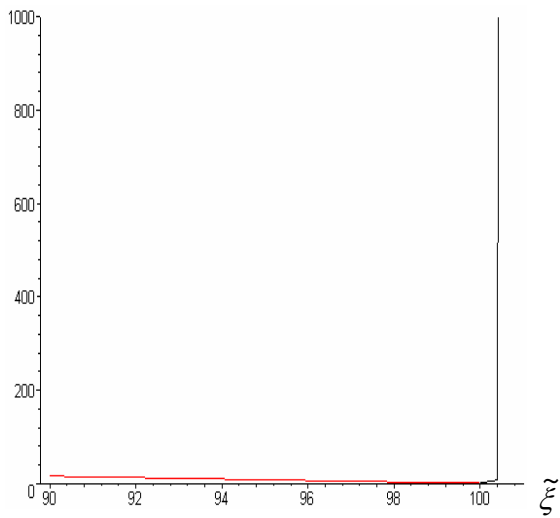
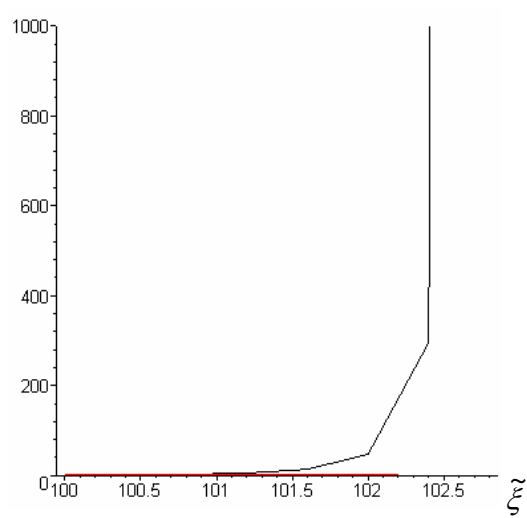

Fig. 4.3. $\tilde{u}(100)=1, \dfrac{\partial \tilde{u}}{\partial \tilde{\xi}}(100)=-1, \tilde{\rho}(100)=1$.   Fig. 4.4. $\tilde{u}(100)=1, \dfrac{\partial \tilde{u}}{\partial \tilde{\xi}}(100)=1, \tilde{\rho}(100)=1$.

All figures contain dependence the dimensionless density $\tilde{\rho}$ and velocity $\tilde{u}$ on $\tilde{\xi}$. In these cases density is practically the vertical line, in chosen scale the curve $\tilde{u}$ is located very close to the abscissa. All blow up solutions demonstrate the existence of the solitons solutions for Eqs. (4.1), (4.2).

**B. GHE solutions in acoustics. Infra-sound as the possible origin of the meteorological dependence of human beings.**

Let us return to the Step 3 and write down the system of equations (3.11) – (3.13) for the generalized Euler equations, where for rarefied gas parameter of non-locality $\tau^{(0)}$ defined by relation (1.9)

(continuity equation)

$$\frac{\partial}{\partial t}\left\{\rho - \tau^{(0)}\left[\frac{\partial \rho}{\partial t} + \frac{\partial}{\partial x}(\rho v_0)\right]\right\} + \frac{\partial}{\partial x}\left\{\rho v_0 - \tau^{(0)}\left[\frac{\partial}{\partial t}(\rho v_0) + \frac{\partial}{\partial x}(\rho v_0^2) + \frac{\partial p}{\partial x}\right]\right\} = 0, \quad (4.17)$$

(motion equation)

$$\frac{\partial}{\partial t}\left\{\rho v_0 - \tau^{(0)}\left[\frac{\partial}{\partial t}(\rho v_0) + \frac{\partial}{\partial x}(\rho v_0^2) + \frac{\partial p}{\partial x}\right]\right\} + \frac{\partial}{\partial x}\left\{\rho v_0^2 + p - \tau^{(0)}\left[\frac{\partial}{\partial t}(\rho v_0^2 + p) + \frac{\partial}{\partial x}(\rho v_0^3 + 3pv_0)\right]\right\} = 0, \quad (4.18)$$

(energy equation)

$$\frac{\partial}{\partial t}\left\{\rho v_0^2 + 3p - \tau^{(0)}\left[\frac{\partial}{\partial t}(\rho v_0^2 + 3p) + \frac{\partial}{\partial x}(\rho v_0^3 + 5pv_0)\right]\right\} + \frac{\partial}{\partial x}\left\{\rho v_0^3 + 5pv_0 - \tau^{(0)}\left[\frac{\partial}{\partial t}(\rho v_0^3 + 5pv_0) + \frac{\partial}{\partial x}\left(\rho v_0^4 + 8pv_0^2 + 5\frac{p^2}{\rho}\right)\right]\right\} = 0. \quad (4.19)$$

Upper index for $\tau^{(0)}$ underlines that mean time between the particles collisions in the rarefied gas is calculated for the local Maxwellian distribution function. In the following this upper index will be omitted.

Let us consider the classical non-stationary 1D hydrodynamic problem when the perturbation of density $\rho'$ and oscillating velocity $v$ are small. Moreover following the classical formulation of the problem we suppose that the temperature $T = const$ and energy equation can be removed from the consideration.

Let us remind the classical Navier-Stokes solution for this case. The linearized system of continuity equation and Navier-Stokes equation is ($\rho = \rho_0 + \rho'$)

$$\frac{\partial \rho'}{\partial t} + \rho_0 \frac{\partial v}{\partial x} = 0, \quad (4.20)$$

$$\rho_0 \frac{\partial v}{\partial t} + c^2 \frac{\partial \rho'}{\partial x} - b\frac{\partial^2 v}{\partial x^2} = 0, \quad (4.21)$$

where $b$ is the efficient viscosity coefficient containing the dynamical viscosity $\mu$ and the bulk viscosity $\eta$

$$b = \frac{4}{3}\mu + \eta, \quad (4.22)$$

$c$ is the sound velocity, $c = \sqrt{RT}$; $R$ is the gas constant.

After differentiating in time Eq. (4.21) and eliminating the derivative $\partial \rho'/\partial t$ one obtains

$$\frac{\partial^2 v}{\partial t^2} - c^2 \frac{\partial^2 v}{\partial x^2} - \frac{b}{\rho_0}\frac{\partial^3 v}{\partial x^2 \partial t} = 0. \quad (4.23)$$

The last term of the left-hand-side of Eq. (4.23) is characterizing the wave damping because of viscosity. Eq. (4.23) admits the exact solution as a product

$$v = y(x)w(t),  \qquad (4.24)$$

which leads to the explicit expressions

$$y = C_1 e^{x\sqrt{A}} + C_2 e^{-x\sqrt{A}}  \qquad (4.25)$$

$$w = C_3 \exp\left[\left(ABc + \sqrt{c^2 A^2 B^2 + 4A}\right)\frac{ct}{2}\right] + C_4 \exp\left[\left(ABc - \sqrt{c^2 A^2 B^2 + 4A}\right)\frac{ct}{2}\right], \qquad (4.26)$$

where $B = \dfrac{b}{c^2 \rho_0}$. Constants $A, C_1, C_2, C_3, C_4$ are defined by the initial and boundary conditions.

Constant $B$ has the order of $\tau$ - mean time between collisions. Important remark follows from the structure the (4.24) – (4.26) solution. It is known from acoustics that classical Euler equations can lead to wave sound solutions even the initial perturbation has not the wave character. For this situation the Navier Stokes description, involving viscosity, leads to relaxation of perturbations without formations of sound waves. Let us specify the local velocity perturbation $v(x)$ at the initial time moment. Then from (4.26) follows that in following time moments the dependence $w(t)$ has the exponential character without waves.

Let us show that generalized hydrodynamic equations can lead to the radically another solutions [17]. Particularly new effects can be found by investigation of atmosphere perturbations as result of cyclone and anti-cyclone evolution.

In the previous assumptions the system follows from (4.17), (4.18)
(continuity equation)

$$\frac{\partial \rho'}{\partial t} - \frac{\partial}{\partial t}\left\{\tau\left[\frac{\partial \rho'}{\partial t} + \rho_0 \frac{\partial v}{\partial x}\right]\right\} + \rho_0 \frac{\partial v}{\partial x} - \left\{\tau\left[\rho_0 \frac{\partial v}{\partial t} + c^2 \frac{\partial \rho'}{\partial x}\right]\right\} = 0$$

(motion equation)

$$\rho_0 \frac{\partial v}{\partial t} - \frac{\partial}{\partial t}\left\{\tau\left[\rho_0 \frac{\partial v}{\partial t} + c^2 \frac{\partial \rho'}{\partial x}\right]\right\} + c^2 \frac{\partial \rho'}{\partial x} - \frac{\partial}{\partial x}\left\{\tau\left[c^2 \frac{\partial \rho'}{\partial t} + 3 p_0 \frac{\partial v}{\partial x}\right]\right\} = 0, \qquad (4.27)$$

In the system (4.27) $\tau$ (which depends in the hydrodynamic approximation on dynamic viscosity and pressure) appears in equations under the derivative sign. As result in the first approximation $\tau$ can be considered as the constant value. Really

$$\frac{\partial \tau}{\partial t} = \Pi\mu \frac{\partial}{\partial t}\left(\frac{1}{p}\right) = -\frac{\Pi\mu}{c^2 \rho_0^2}\frac{\partial \rho'}{\partial t}. \qquad (4.28)$$

As we see the derivatives $\partial\tau/\partial t, \partial\tau/\partial x$ are the values of the first order and after substitution in the system (4.27) could lead to squared perturbations which are omitted. As result we have from (4.27):
(continuity equation)

$$\frac{\partial \rho'}{\partial t} - \tau \frac{\partial^2 \rho'}{\partial t^2} - 2\rho_0 \tau \frac{\partial^2 v}{\partial t \partial x} + \rho_0 \frac{\partial v}{\partial x} - \tau c^2 \frac{\partial^2 \rho'}{\partial x^2} = 0, \qquad (4.29)$$

(motion equation)

$$\rho_0 \frac{\partial v}{\partial t} - \tau \rho_0 \frac{\partial^2 v}{\partial t^2} - 2\tau c^2 \frac{\partial^2 \rho'}{\partial x \partial t} + c^2 \frac{\partial \rho'}{\partial x} - 3\tau p_0 \frac{\partial^2 v}{\partial x^2} = 0, \qquad (4.30)$$

After differentiating Eq. (4.30) in time

$$\rho_0 \frac{\partial^2 v}{\partial t^2} - \tau\rho_0 \frac{\partial^3 v}{\partial t^3} - 2\tau c^2 \frac{\partial^2}{\partial x \partial t}\frac{\partial \rho'}{\partial t} + c^2 \frac{\partial}{\partial x}\frac{\partial \rho'}{\partial t} - 3\tau p_0 \frac{\partial^3 v}{\partial x^2 \partial t} = 0 \quad (4.31)$$

we use the continuity equation

$$\frac{\partial}{\partial t}\left(\rho' - \tau \frac{\partial \rho'}{\partial t}\right) - \tau c^2 \frac{\partial^2 \rho'}{\partial x^2} = 2\rho_0 \tau \frac{\partial^2 v}{\partial t \partial x} - \rho_0 \frac{\partial v}{\partial x}, \quad (4.32)$$

for eliminating the derivative $\partial \rho'/\partial t$, one obtains

$$\tau \frac{\partial^3 v}{\partial t^3} - \frac{\partial^2 v}{\partial t^2} - \tau c^2 \frac{\partial^3 v}{\partial x^2 \partial t} + c^2 \frac{\partial^2 v}{\partial x^2} = 0 \ . \quad (4.33)$$

It should be noticed that the following estimation was used

$$\frac{\partial \rho'}{\partial t} \approx 2\rho_0 \tau \frac{\partial^2 v}{\partial t \partial x} - \rho_0 \frac{\partial v}{\partial x}, \quad (4.34)$$

by the derivation (4.33) and the relations $p_0 = \rho_0 RT = \rho_0 c^2$, $\tau = \Pi \frac{\mu}{\rho_0 c^2}$.

Let us compare Eq. (4.33) with its Navier-Stokes analogue (4.23). Eq. (4.33) has the higher order in time and contains the time derivative of the third time. This term can not be omitted if the parameter $\tau$ is not small, but this term cannot be omitted also for small $\tau$ as the term containing the small parameter in front of the senior derivative. Eq. (4.34) contains also cross-derivative space-time as Eq. (4.23) but with another sign.

The solution of the Eq. (4.33) is written as

$$v = F_1(x - ct) + F_2(x + ct), \quad (4.35)$$

where $F_1(x - ct)$, $F_2(x + ct)$ - a wave functions of arguments $(x - ct)$ and $(x + ct)$. Explicit form of these functions is defined by the initial and boundary conditions.

Therefore in the formulated assumptions GHE lead to two waves propagating in the viscous medium along positive and negative directions of $x$ without damping with the sound velocity. Modern software like Maple 10 can obtain the analytical solutions of linear differential equations (4.29), (4.30), but these solutions has rather huge form and better to use the numerical approach for the system solution. Numerical solution of the non-linear generalized hydrodynamic equations (GHE) confirms the existence of infra-sound waves after the meteorological fronts corresponding to evolution cyclones ant anti-cyclones in atmosphere.

This extremely important fact can explain so called meteorological dependence of the human beings.

**Conclusion.**
As it is shown the theory of transport processes (including quantum mechanics) can be considered in the frame of unified theory based on the non-local physical description. In particular the generalized hydrodynamic equations represent an effective tool for solving problems in the very vast area of physical problems. The results of this investigation are published in [18].

**Acknowledgments**
Author is thankful to Dr. A. Fedoseyev and Dr. I. Ovchinnikova for fruitful discussions.

**Appendix 1. Perturbation method of the equation solution related to $T[f]$.**

Let us consider the solution of Eq (2.32) by the perturbation method. Equation (2.32) is valid only by the small $y$ and moreover the function $T[f]$ receives the physical meaning only by $y = 0$. Let us expand $T[f]$ in series

$$T[f](t,x,y) = \sum_{n=0}^{\infty} T_n[f](t,x) y^n. \qquad (A.1)$$

After substitution (A.1) in equation (2.32)

$$i\alpha \frac{\partial}{\partial t} T[f](x,y,t) = -\frac{\alpha^2}{2m} \frac{\partial^2}{\partial x \partial y} T[f](x,y,t) - 2yF(x)T[f](x,y,t).$$

we find after equalizing the terms in front of the same powers of $y$;

for $y^0$

$$i \frac{\partial}{\partial t} T_0[f](x,t) + \frac{\alpha}{2m} \frac{\partial}{\partial x} T_1[f](x,t) = 0. \qquad (A.2)$$

for $y^1$

$$i\alpha \frac{\partial}{\partial t} T_1[f](x,t) = -\frac{\alpha^2}{m} \frac{\partial}{\partial x} T_2[f](x,t) - 2F(x)T_0[f](x,t). \qquad (A.3)$$

for $y^2$

$$i\alpha \frac{\partial}{\partial t} T_2[f](x,t) = -\frac{3\alpha^2}{2m} \frac{\partial}{\partial x} T_3[f](x,t) - 2F(x)T_1[f](x,t). \qquad (A.4)$$

As we see the successive approximations lead to the chain of equations and the first links of the chain are Eqs. (A.2) – (A.4).

The coefficients of the expansion (A.1) are, generally speaking, the complex functions with one exception concerning to the coefficient $T_0[f](x,t)$ because of the condition

$$T[f](x, y = 0, t) = T_0[f](x,t) = |\Psi(x,t)|^2$$

Denoting $\rho$ as real probability density ($|\Psi(x,t)|^2 = \rho$), one obtains from (A.2)

$$i \frac{\partial}{\partial t} \rho(x,t) + \frac{\alpha}{2m} \frac{\partial}{\partial x} T_1[f](x,t) = 0. \qquad (A.5)$$

But (see. (2.28))

$$T[f](x,y,t) = \Psi^*(t, x-y) \Psi(t, x+y),$$

and

$$\Psi^*(t, x-y) = \sum_{n=0}^{\infty} \Psi_n^*(t,x)(-y)^n = \Psi_0^*(t,x) - \left(\frac{\partial \Psi_0^*}{\partial x}\right)_{y=0} y + \frac{1}{2}\left(\frac{\partial^2 \Psi_0^*}{\partial x^2}\right)_{y=0} y^2 - \ldots, \qquad (A.6)$$

$$\Psi(t, x+y) = \sum_{n=0}^{\infty} \Psi_n^*(t,x) y^n = \Psi_0(t,x) + \left(\frac{\partial \Psi_0}{\partial x}\right)_{y=0} y + \frac{1}{2}\left(\frac{\partial^2 \Psi_0}{\partial x^2}\right)_{y=0} y^2 + \ldots, \qquad (A.7)$$

then

$$T_1(x,t) = \Psi_0^*(t,x)\left(\frac{\partial \Psi_0}{\partial x}\right)_{y=0} - \Psi_0(t,x)\left(\frac{\partial \Psi_0^*}{\partial x}\right)_{y=0}. \qquad (A.8)$$

From (A.8) follows that $T_1(x,t)$ is an imagine value. Really,

$$T_1(x,t) = \Psi_0^*(t,x)\left(\frac{\partial \Psi_0}{\partial x}\right)_{y=0} - \Psi_0(t,x)\left(\frac{\partial \Psi_0^*}{\partial x}\right)_{y=0} = \Psi_0^*(t,x)\left(\frac{\partial \Psi_0}{\partial x}\right)_{y=0} - \left[\Psi_0^*(t,x)\left(\frac{\partial \Psi_0}{\partial x}\right)_{y=0}\right]^*.$$

This result coincides with Eq. (A.5), from which follows

$$\frac{\partial}{\partial x} T_1^{real}[f](x,t) = 0, \quad (A.10)$$

or

$$T_1^{real}[f](x,t) = const. \quad (A.11)$$

Then, $const = 0$ in (A.11) and Eq. (A.5) takes the form

$$\frac{\partial}{\partial t}\rho(x,t) + \frac{\alpha}{2m}\frac{\partial}{\partial x} T_1^{imagine}[f](x,t) = 0, \quad (A.12)$$

because

$$T_1[f](x,t) = i T_1^{imagine}[f](x,t) \quad (A.13)$$

From Eqs. (A.5), (A.8) follow

$$i\frac{\partial}{\partial t}\rho(x,t) + \frac{\alpha}{2m}\frac{\partial}{\partial x}\left[\Psi_0^*(t,x)\left(\frac{\partial \Psi_0}{\partial x}\right)_{y=0} - \Psi_0(t,x)\left(\frac{\partial \Psi_0^*}{\partial x}\right)_{y=0}\right] = 0 \quad (A.14)$$

or

$$i\Psi_0(t,x)\frac{\partial}{\partial t}\Psi_0^*(t,x) + i\Psi_0^*(t,x)\frac{\partial}{\partial t}\Psi_0(t,x) + \frac{\alpha}{2m}\left[\Psi_0^*(t,x)\left(\frac{\partial^2 \Psi_0}{\partial x^2}\right)_{y=0} - \Psi_0(t,x)\left(\frac{\partial^2 \Psi_0^*}{\partial x^2}\right)_{y=0}\right] = 0$$

(A.15)

Eq. (A.15) is satisfied identically if

$$i\frac{\partial}{\partial t}\Psi_0(t,x) + \frac{\alpha}{2m}\left(\frac{\partial^2 \Psi_0}{\partial x^2}\right)_{y=0} = 0. \quad (A.16)$$

Eq. (A.16) is Schrödinger equation of the first approximation.
For the second approximation (see (A.3), (A.8))

$$i\alpha\frac{\partial}{\partial t}\left[\Psi_0^* \frac{\partial \Psi_0}{\partial x} - \Psi_0 \frac{\partial \Psi_0^*}{\partial x}\right] = -\frac{\alpha^2}{m}\frac{\partial}{\partial x} T_2[f](x,t) - 2F(x)T_0[f](x,t), \quad (A.17)$$

and

$$T_2(x,t) = \frac{1}{2}\Psi_0^*(t,x)\left(\frac{\partial^2 \Psi_0}{\partial x^2}\right)_{y=0} + \frac{1}{2}\Psi_0(t,x)\left(\frac{\partial^2 \Psi_0^*}{\partial x^2}\right)_{y=0} - \left(\frac{\partial \Psi_0}{\partial x}\right)_{y=0}\left(\frac{\partial \Psi_0^*}{\partial x}\right)_{y=0}. \quad (A.18)$$

After substitution (A.18) in Eq.(A.17)

$$i\alpha\frac{\partial}{\partial t}\left[\Psi_0^*\frac{\partial \Psi_0}{\partial x} - \Psi_0\frac{\partial \Psi_0^*}{\partial x}\right] = -\frac{\alpha^2}{2m}\frac{\partial}{\partial x}\left[\Psi_0^*\frac{\partial^2 \Psi_0}{\partial x^2} + \Psi_0\frac{\partial^2 \Psi_0^*}{\partial x^2} - 2\frac{\partial \Psi_0}{\partial x}\frac{\partial \Psi_0^*}{\partial x}\right] - 2F(x)\Psi_0^*\Psi_0$$

(A.19)

Let us consider the equation

$$i\alpha\frac{\partial}{\partial t}\left[\Psi_0^*\frac{\partial \Psi_0}{\partial x}\right] = -\frac{\alpha^2}{2m}\frac{\partial}{\partial x}\left[\Psi_0^*\frac{\partial^2 \Psi_0}{\partial x^2} - \frac{\partial \Psi_0}{\partial x}\frac{\partial \Psi_0^*}{\partial x}\right] - F(x)\Psi_0^*\Psi_0. \quad (A.20)$$

The conjugate equation can be written as

$$-i\alpha\frac{\partial}{\partial t}\left[\Psi_0\frac{\partial \Psi_0^*}{\partial x}\right] = -\frac{\alpha^2}{2m}\frac{\partial}{\partial x}\left[\Psi_0\frac{\partial^2 \Psi_0^*}{\partial x^2} - \frac{\partial \Psi_0^*}{\partial x}\frac{\partial \Psi_0}{\partial x}\right] - F(x)\Psi_0^*\Psi_0. \quad (A.21)$$

Summation of equations (A.20), (A.21) gives Eq. (A.19). From Eq. (A.20) follows equation which could be titled as Schrödinger equation of the second approximation

$$i\alpha \frac{\partial \Psi_0^*}{\partial t}\frac{\partial \Psi_0}{\partial x} + i\alpha \Psi_0^* \frac{\partial^2 \Psi_0}{\partial t \partial x} = -\frac{\alpha^2}{2m}\left[\Psi_0^* \frac{\partial^3 \Psi_0}{\partial x^3} - \frac{\partial \Psi_0}{\partial x}\frac{\partial^2 \Psi_0^*}{\partial x^2}\right] - F(x)\Psi_0 \Psi_0^* \qquad (A.22)$$

Eq. (A.22) can be transformed as

$$i\alpha \frac{\partial \ln \Psi_0^*}{\partial t}\frac{\partial \Psi_0}{\partial x} + i\alpha \frac{\partial^2 \Psi_0}{\partial t \partial x} = -\frac{\alpha^2}{2m}\left[\frac{\partial^3 \Psi_0}{\partial x^3} - \frac{1}{2\rho}\frac{\partial \Psi_0^2}{\partial x}\frac{\partial^2 \Psi_0^*}{\partial x^2}\right] - F(x)\Psi_0 . \qquad (A.23)$$

As we see Schrödinger equation of the second approximation is non-linear equation of the third order in space, containing the cross derivative "time-space" and derivative on time of the logarithmic term.

REFERENCES
[1] L. Boltzmann, Weitere Studien über das Wärmegleichgewicht unter Gasmolekulen, Sitz. Ber. Kaiserl. Akad. Wiss. Band 66(2) 275 (**1872**)
[2] L. Boltzmann, Vorlesungen über Gastheorie, Verlag von Johann Barth, Leipzig (**1912**).
[3] S. Chapman and T.G. Cowling, The Mathematical Theory of Non-uniform Gases, At the University Press, Cambridge (**1952**)
[4] I. O. Hirschfelder, Ch. F. Curtiss and R.B. Bird, Molecular Theory of Gases and Liquids, John Wiley and sons, inc. New York. Chapman and Hall, lim., London (**1954**)
[5] B.V., Alekseev, Matematicheskaya Kinetika Reagiruyushchikh Gazov (Mathematical Theory of Reacting Gases), Nauka, Moscow (**1982**)
[6] B.V. Alexeev, The Generalized Boltzmann Equation, Generalized Hydrodynamic Equations and their Applications, Phil. Trans. Roy. Soc. Lond. vol.349, 417 (**1994**)
[7] B.V. Alexeev, The Generalized Boltzmann Equation, Physica A. vol. 216, 459 (**1995**)
[8] B.V. Alekseev, Physical Principles of the Generalized Boltzmann Kinetic Theory of Gases" Physics-Uspekhi. vol.43 (6), 601 (**2000**)
[9] B.V. Alekseev, Physical Fundamentals of the Generalized Boltzmann Kinetic Theory of Ionized Gases. Physics-Uspekhi, vol. 46 (2), 139 (**2003**)
[10] B.V. Alexeev Generalized Boltzmann Physical Kinetics. Elsevier (**2004**)
[11] Yu. L. Klimontovich., About Necessity and Possibility of Unified Description of Hydrodynamic Processes. Theoretical and Math. Physics, vol. 92 (2) 312 (**1992**)
[12] J.S. Bell, On the Einstein Podolsky Rosen Paradox, Physics, vol. 1,195 (**1964**)
[13] A.I. Fedoseyev, M. Turowski and M.S. Wartak, Kinetic and Quantum Models for Nanoelectronic and Optoelectronic Device Simulation. Journal of Nanoelectronics and Optoelectronics, vol. 2, 3, 234 (**2007**)
[14] B.N. Rodimov, Auto-Oscillating Quantum Mechanics, Tomsk, Tomsk State University Publ. (**1976**).
[15] B.V. Alexeev, A.I. Abakumov and V.S. Vinogradov, Mathematical Modeling of Elastic Interactions of Fast Electrons with Atoms and Molecules. Communications on the Applied Mathematics. Computer Centre of the USSR Academy of Sciences, Moscow (**1986**)
[16] E. Carnovalli Jr, H.M. Franca. On the Connection Between the Liouville Equation and the Schrödinger Equation. Arxiv, quant-ph/0512049 vol. 2, 17 (**2006**}
[17] B.V. Alexeev, Application of the Generalized Hydrodynamic Equations in Acoustics. Infrasound as the Possible Cause of the Meteorological Dependence of Human Beings. Ecological systems and sets. vol. 5, 33 (**2006**)
[18] B.V. Alexeev, Generalized Quantum Hydrodynamics and Principles of Non-Local Physics, Journal of Nanoelectronics and Optoelectronics. vol. 3, 1 – 16 (**2008**)